*Review*

# The Blue Supergiant Problem and the Main-Sequence Width

**Jorick S. Vink [1,2,*] and Rene D. Oudmaijer [2,3]**

[1] Armagh Observatory and Planetarium, College Hill, Armagh BT61 9DG, UK
[2] School of Physics & Astronomy, University of Leeds, Woodhouse Lane, Leeds LS2 9JT, UK; r.d.oudmaijer@leeds.ac.uk
[3] Royal Observatory of Belgium, Ringlaan 3, 1180 Brussels, Belgium
* Correspondence: jorick.vink@armagh.ac.uk

**Abstract:** Using Gaia DR3 we derive new distances and luminosities for a sample of Galactic B supergiants which were thought to be post main-sequence (MS) objects from their HR diagram location beyond the terminal-age MS (TAMS). When applying the newer Gaia distances in addition to enhanced amounts of core-boundary mixing, aka convective overshooting, we show that these Galactic B supergiants are likely enclosed within the MS band, indicating an evolutionary stage of steady core hydrogen burning. We discuss the importance of considering enhanced overshooting and how vectors in the mass-luminosity plane (ML-plane) can be used to disentangle the effects of wind mass loss from interior mixing. We finish with the key message that any proposed solution to the BSG problem should consider not only an explanation for the sheer number of B supergiants inside the Hertzsprung gap, but should at the same time also account for the steep drop in rotation rates identified at spectral type B1—corresponding to an effective temperature of ∼21 kK, and for which two distinct families of solutions have been proposed.

## 1. Introduction on the Blue Supergiant Problem

The question on the origin of B[e] stars in general and that of B[e] supergiants more specifically should probably not be tackled in isolation but as part of the wider puzzle of the origin of canonical B supergiants[1]. While researchers in the supernova community often discuss an issue dubbed the red supergiant (RSG) problem [1,2], there is in fact no real problem explaining the location nor the quantity of RSGs in the HR diagram from standard evolutionary models [3,4]. Instead, there is a blue supergiant (BSG) problem, as there are far too many B supergiants located inside the Hertzsprung gap of the stellar HR diagram [5–8], where stars are supposed to rapidly traverse through on their way to becoming RSGs.

In addition to canonical B supergiants, the zoo of objects in the blue part of the HR diagram also includes objects surrounded by dusty circumstellar media, such as B[e] supergiants and S Dor luminous blue variables (LBVs). While the physics of rapid rotation probably plays the key role in the formation scenario of outflowing disks around B[e] supergiants [9], it is physics related to the Eddington limit that likely plays a decisive role in creating S Dor LBV variability [10]. In some cases, such as for the famous LBV AG Car, both S Dor LBV variability as well as B[e] disk characteristics are present, which may possibly require a physical explanation that involves both the Eddington limit as well as stellar rotation [11].

One scenario that has become popular is that involving stellar merging to explain both of these B[e] and LBV phenomena [12–15]. Perhaps surprisingly, binary mergers have more recently also been invoked to explain the general BSG problem [16–18]. While stellar mergers might be an attractive way to explain the rapid rotation of B[e] supergiants, as well as the explodibility of LBV supernovae [15], mergers could also lead to the formation of a magnetic field, which could brake the star—leading to slow rotation [19].

It is however somewhat uncomfortable that binary mergers have been involved to explain very *rapid* and very *slow* stellar rotation simultaneously, and it is pertinent that we try to understand the general characteristics of the B supergiant population, before accepting any one particular explanation for the BSG problem. There are two key factors of the BSG problem that are sometimes overlooked even in recent literature, that is: (i) the sheer number of B supergiants present in the HR diagram and (ii) the fact that B supergiants are slow rotators below a $T_{\rm eff}$ of 21 kK (see [20]and the right-hand side of the plot at the end of this manuscript).

Stellar expansion from the main sequence is often employed to explain the slower rotation of B supergiants compared to their O star predecessors [21], but this seems to overlook the fact that after the MS, the stellar lifetimes are so much shorter that the B supergiants should not be abundantly present in the first place. Instead, stellar mergers with a B-field might explain the slow rotation as well as the large number of B supergiants [16] but does not explain the key feature of the steep $v \sin i$ drop at 21 kK.

Crowther et al. [22] studied a sample of Galactic B supergiants with the non-LTE and stellar wind code CMFGEN finding nitrogen (N) abundances that were roughly an order of magnitude higher than the solar value. Stellar effective temperatures were accurately derived between 15 and 30 kK, which placed the entire B supergiant sample beyond the terminal-age main sequence (TAMS) with the relatively low overshooting ($\alpha_{\rm ov}$ of about 0.1) stellar models at the time, see Figure 5 in [22]. However, B supergiant luminosities could not be accurately derived due to the lack of Gaia astrometric distances. In addition, there has been a flurry of activity in our understanding of convective-boundary mixing (CBM), aka convective overshooting in the last decade [23].

## 2. The Related Problem of the Main-Sequence Width

Regarding CBM many stellar evolution models adopt the step-overshooting method, increasing the convective core during core H-burning by a fraction $\alpha_{\rm ov}$ of the pressure scale height $H_p$, noting that the extension of the convective core by $\alpha_{\rm ov}$ dredges fresh H from the envelope into the core, replenishing the supply of H fuel and extending the H-burning phase of evolution by tens of per cent, widening the MS.

Martins & Palacios [24] compared various grids of massive star evolutionary models including different chemical mixing (and other) assumptions. It was claimed that the MS width is ever so slightly too narrow in comparison to massive star observations for the low overshooting ($\alpha_{\rm ov} = 0.1$) Geneva models of [25], but far too wide for the moderate overshooting $\alpha_{\rm ov}$ Bonn models of [26]. This conclusion was primarily based on a study of the location of observed stars in the HR diagram. An overshooting parameter $\alpha_{\rm ov}$ between 0.1 and 0.2 in Geneva models with rotation was preferred to reproduce the main sequence width of massive stars (see also the more recent Geneva models of [21]). However, the conclusion that the low core overshooting ($\alpha = 0.1$) models agree better with observations is largely based on the *assumption* that only O-type dwarfs of luminosity class IV and V are main sequence objects, and that both O supergiants and B supergiants would automatically be located beyond the TAMS, is rather questionable. The issue with such arguments is that they are somewhat circular as they inherently assume that only dwarfs can be core H burning. While an evolutionary distinction between dwarfs and supergiants is indeed applicable to low-mass stars, there is no reason for this also to be the case for high-mass stars with their large convective cores. Simply put, there is little reason why a stellar atmosphere, with spectral luminosity class classification related to atmospheric $\log g$, would somehow know the nuclear burning stage inside the stellar core.

In other words, the simple usage of luminosity classes and HR diagram location alone are not sufficient to resolve the BSG puzzle, and alternative diagnostics need to be considered. One such method is the steep drop of rotational velocities at an effective temperature of 21 kK. O-type stars on the hot side of this jump show both slow and rapid stellar rotation, while B supergiants on the cool side are *all* slow rotators [20]. This feature was not only present in the Large Magellanic Cloud (LMC) spectra of the VLT-Flames survey [27] but also in Galactic OB supergiant rotation data of [28], first plotted in Vink et al. [20].

We proposed two possible interpretations of this steep $v \sin i$ drop. One was that the feature is related to the TAMS (see also [29], the second one was that it could be connected to an increased amount of wind mass loss at the bi-stability jump located at $T_{\rm eff}$ = 21,000 K[2]. Both possible explanations require a higher amount of core overshooting than the small amount of order $\alpha_{\rm ov}$ = 0–0.1 [32,33] employed at the time. In the case the rotational drop feature represents the TAMS, $\alpha_{\rm ov}$ could be of order 0.3 in line with [26], and in case it is related to the bi-stability jump the number $\alpha_{\rm ov}$ = 0.3 would just be a lower limit, and the real value could be of order 0.4–0.5 [20,34].

## 3. Data & Method

In this work we provide updated distances and luminosities of the Galactic B supergiant sample of [22] which has been studied with non-LTE atmospheres giving reliable nitrogen and carbon abundances. While the temperatures spanning a wide range between 15 and 30 kK were reliable due to the line-blanketing physics included in these nLTE models, the luminosities were uncertain due to unreliable distances at the time. Here we redetermine the luminosities of these B supergiants using the revised distances made possible by Gaia DR3 [35], in order to provide an accurate comparison in the HR diagram to stellar models (see Figure 1).

Gaia has obtained astrometric data for almost 2 billion objects, which include parallaxes to a large precision. These parallaxes were converted into geometric distances by [36], who used a standard model for the Galaxy as a prior[3] to correct for possible selection biases and the zero-point offset. The luminosities were then re-scaled using the new distance and the distance published in the datatables in the respective papers, a method previously applied by e.g., Oudmaijer et al. [38] who presented new luminosities for lower mass post-Asymptotic Giant Branch stars.

We also plot non-rotating Galactic stellar models from [34] which were calculated using the `MESA` stellar evolution code for a range of initial masses from 8 to 60 $M_\odot$. For CBM these models adopt the step-overshooting $\alpha_{\rm ov}$ with values of $\alpha_{\rm ov}$ = 0.1 and 0.5. As can be noted from Figure 1 for the higher assumed values of overshoot all B supergiants in this temperature range could be explained as still burning H in the stellar core. Note that this does not imply that *all* BSGs are indeed core H burning. In fact there are also cooler B supergiants at later spectral types below 15 kK, such as those analysed by [39] that would be located on the cool side of the high overshooting $\alpha_{\rm ov}$ = 0.5 TAMS, allowing later B supergiants to be post-MS. A similar conclusion was reached from the VLT-Flames Tarantula survey (VFTS) study in [40] for B supergiants at roughly 50% $Z_\odot$.

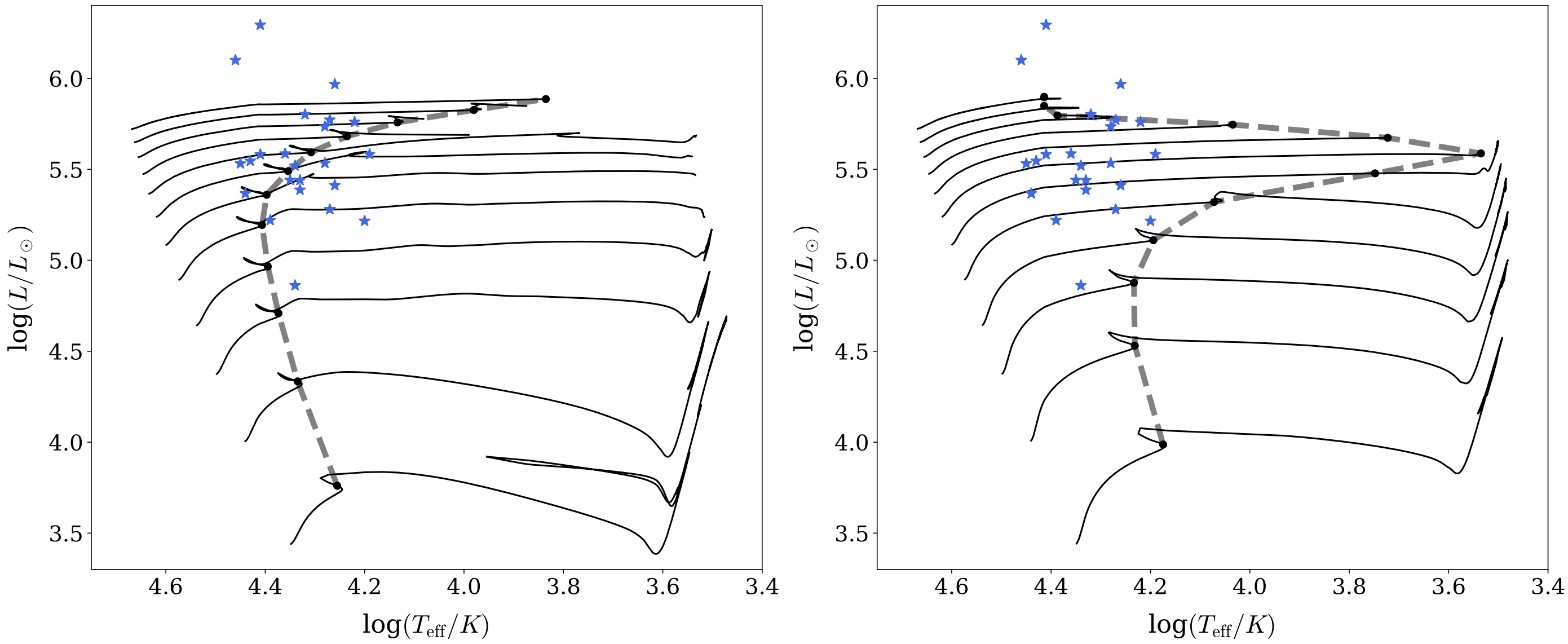

**Figure 1.** HR diagram comparison of Galactic B supergiants with solar metallicity [34] MESA model tracks with initial masses from 8–60 $M_\odot$, in steps 8, 12, 15, 20, 25, 30, 35, 40, 45, 50, 55, 60 $M_\odot$. The figure on the left-hand side is for a small amount of overshooting with $\alpha_{\rm ov} = 0.1$, while the figure on the right-hand side is for $\alpha_{\rm ov} = 0.5$. The dashed lines on both the left and right-hand side denote the TAMS location. The blue stars represent the observed B supergiant sample from Crowther et al. [22] with our updated luminosities utilising Gaia DR3 distances.

So while it is obvious that the lower the temperature of a star in question the higher is the chance it is no longer core H burning MS object, the HR diagram location is not definitive as stellar models rely on a free parameter called convective overshoot.

## 4. Mass-Luminosity Plane: Disentangling Mixing and Mass Loss

In order to make progress in our understanding of stellar evolution, and solving problems such as the BSG problem, it is pertinent that we are able to disentangle the key ingredients that influence massive star evolution, such as wind mass loss and interior mixing.

While asteroseismology could potentially constrain the core sizes of stars, up to today results have only been obtained up to a mass of around 20 $M_\odot$, delivering a rather wide range in observed $\alpha_{\rm ov}$ from 0–0.4 [41] (see Table 1). For more massive stars other methods such as the study of eclipsing binaries becomes relevant, see [42,43]. Still, only using traditional tools, such as the stellar HR diagram is not sufficient to disentangle the various physical ingredients. Higgins & Vink [34] have shown how vectors in the mass-luminosity plane, with mixing elongating the vector, and mass loss changing the vector's angle, see Figure 2 could be used instead.

Employing the newly developed ML-plane method, Higgins & Vink [34] found that in order to reproduce the evolution of one of the stars in the massive detached eclipsing binary HD 166734 that $\alpha_{\rm ov} = 0.5$ was required. Similarly, in [44] the LMC detached eclipsing binary VFTS 500 which lies close to the TAMS required the same extension of $\alpha_{\rm ov} = 0.5$ in order to reproduce the evolution of both components assuming the stars formed at the same time.

This shows on the one hand that significantly enhanced overshooting is mandatory for at least some massive stars, but one should on the other hand also be mindful of the fact that *not all* massive stars studied via asteroseismology or the ML-plane method have such high values of overshoot. Phrased differently, per mass bin, there appears to be a distribution of derived values of $\alpha_{\rm ov}$ over the full OB star range [41,45].

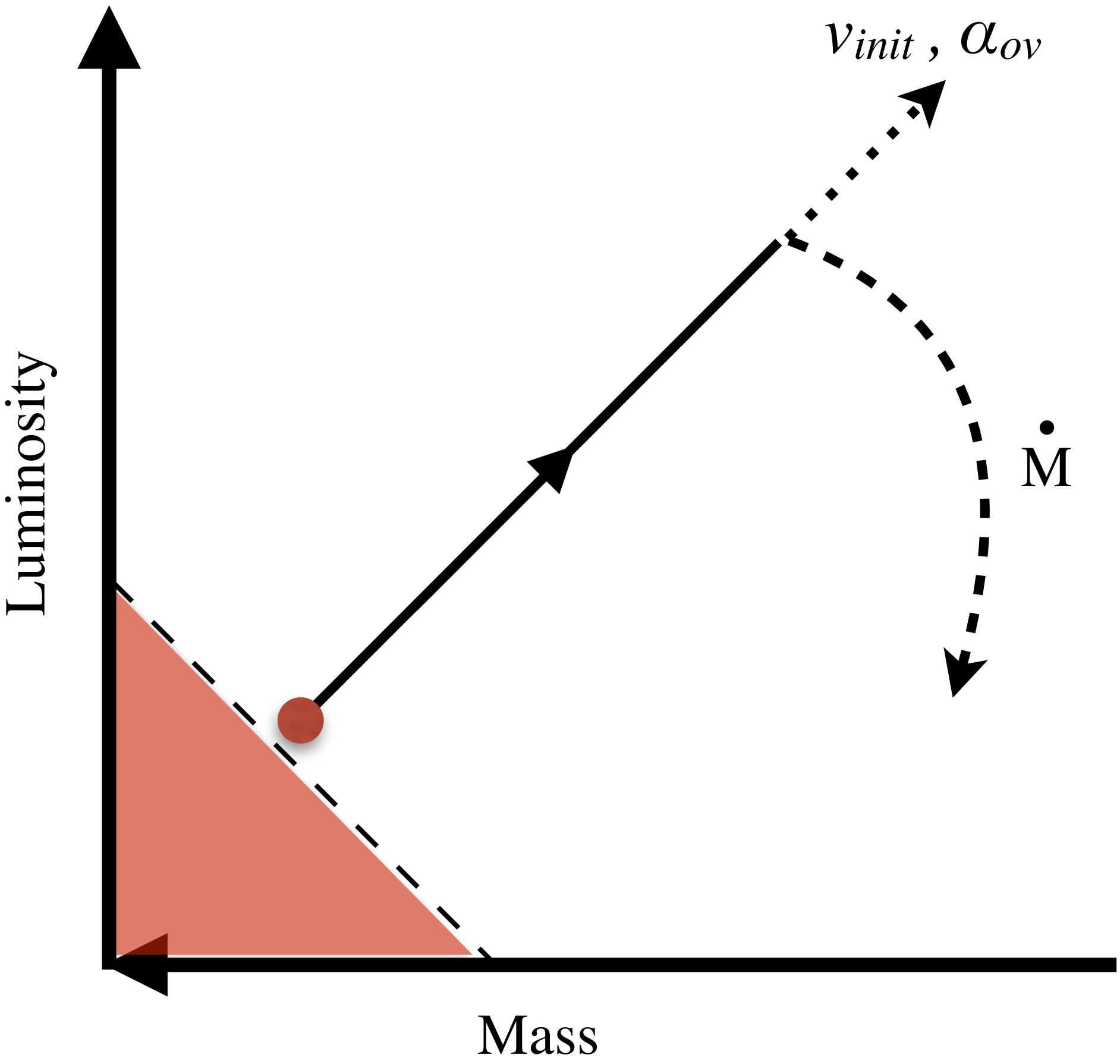

**Figure 2.** An illustration of the mass-luminosity plane, showing a typical evolutionary track that begins at the ZAMS, indicated by the red dot, and progresses along the black arrow towards the TAMS. The dotted vector indicates how factors such as increased rotation and/or convective overshooting can extend the M-L vector. The curved dashed line represents the gradient where mass-loss rates influence this M-L vector. The red solid area is prohibited, as set by the mass-luminosity relationship. Adapted from Higgins & Vink [34].

Sofar we have assumed that a large core is the result of a higher amount of CBM, but this is not necessarily the case. A larger core can in principle be achieved by any kind of extra mixing, such as rotational mixing, either in a single star or in a binary. Additional diagnostics, such as N enhancement [46,47] a function of $v \sin i$ could be utilised to distinguish the contribution to the mixing vector in the ML-plane between for instance rotational mixing versus overshooting [34].

## 5. The Importance of Homogeneous Samples and More Diagnostics

The simple picture that massive stars spend most of their time (90% as core H burning objects) at hot temperatures above 30 kK, before traversing the HR diagram, creating a gap, before arriving at the location of core He burning either as cool RSGs or warmer He-BSGs around 10 kK is simply not tenable.

Castro et al. [48] performed a literature study of a large sample of 600 Galactic OB supergiants finding a gap, which they interpreted as the location of the TAMS in the Milky Way. The issue with that conclusion is that the sample was inherently biased, as it relied on which stars spectroscopic analyses had been performed according to various author's preferences.

Indeed, as is shown on the left-hand side of Figure 3 the unbiased samples of the VLT-Flames and VFTS surveys in the LMC did *not* show a gap in the HR diagram [20]. This thus suggests that the "gap" identified in the Galactic literature study might not be real.

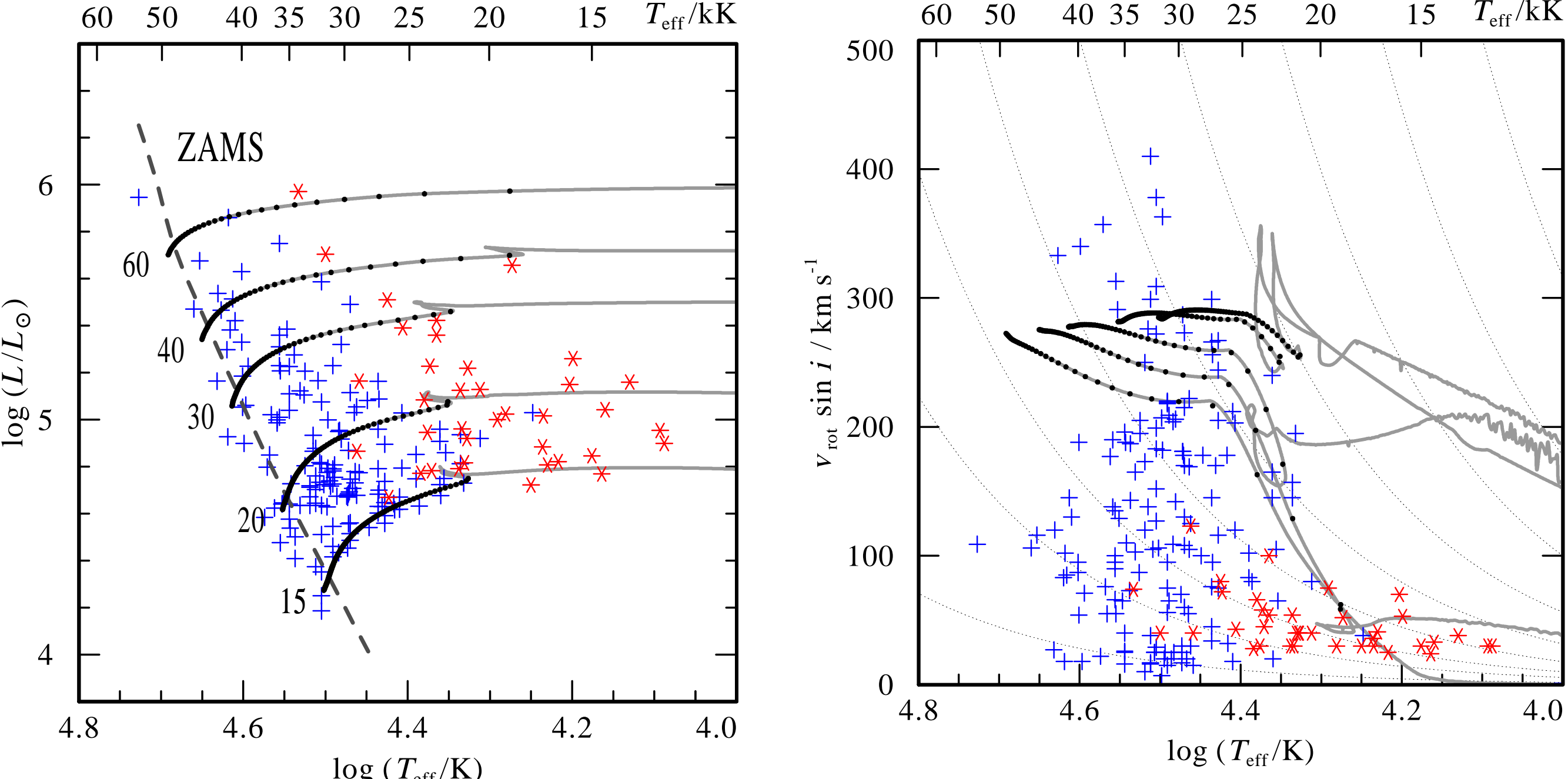

**Figure 3.** Luminosities (Left-hand side) and rotational velocities (right-hand side) versus effective temperature from the VLT Flames survey of massive stars for evolutionary masses over 15 $M_\odot$. Luminosity classes are represented by blue plus signs (for luminosity classes II–V) and red stars (for luminosity class I). The evolutionary tracks for LMC metallicity (50% solar), including the predicted bi-stability jump, are shown in grey, with initial rotation velocities of 250 km/s for five masses: 15, 20, 30, 40, and 60 $M_\odot$. Note that the critical mass for bi-stability braking is approximately 35 $M_\odot$ at LMC metallicity in these specific models computed with the BONN stellar evolution code for a step overshooting value of $\alpha_{\rm ov}$ = 0.5. The rotational velocity tracks can be compared to the angular momentum conservation case, depicted as grey dotted background lines, showing that bi-stability braking can be steeper than would be attributable to angular momentum conservation. The black dots along the evolutionary tracks correspond to $10^5$ year time-steps. Adapted from Vink et al. [20].

## 6. Final Words

As the BSG problem is unresolved, and HR diagram location alone is shown to be insufficient to resolve it, we argue it is pertinent to not only perform stellar evolutionary and future population synthesis studies of the positions of single and binary stars in the HR diagram, but also their rotation rates shown on the right-hand side of Figure 3.

The right-hand side not only shows data-points but also evolutionary tracks from [26] that include a bi-stability jump, which could potentially cause the rapid drop in rotation rates at spectral type B1 above a certain critical mass, which is 35 $M_\odot$ in these particular models at this LMC metallicity. As shown in [20] if the bi-stability jump is not included in the evolutionary models, none of the models would spin down, as efficient angular momentum transport—via a B-field—has been applied in these [26] models.

The main challenge of the bi-stability braking model is that there is still no empirical evidence for a mass loss bi-stability jump [20,31]. The second potential explanation of the feature would be that the drop feature delineates a demarcation of two separate populations. If CBM via $\alpha_{\rm ov}$ was a well-understood phenomenon, this could be the more likely explanation, but it has significant challenges in its own right. As discussed in Section 4 asteroseismic observations in the B-star regime as well as ML-plane O star analyses have shown a wide *distribution* in $\alpha_{ov}$ values between zero and 0.5. i.e., there is no universal TAMS for each stellar mass. There is a second challenge and that is that in order for the TAMS to explain the abrupt 21 kK feature one would need to employ a *decreasing* amount of core overshoot with mass, which is opposite to that predicted by CBM models of [49].

The point is that a *constant* value of $\alpha_{ov}$ would result in a redwards bending as observed in Figure 1, where the TAMS location shifts to cooler temperatures for increasing stellar mass. The reason for the redwards bending is the inflation of the outer envelope [50–53] kicking in at the higher stellar masses. This means that the only way to obtain a TAMS at constant Teff of 21 kK one would not only require a *mass-dependent decrease* of the core overshoot value $\alpha_{ov}$, but also impose a delicate balance between overshoot parametrisation and envelope inflation. This seems extremely unlikely.

So at the moment, both families of explanations for the steep drop in [20] have difficulties explaining the BSG problem: there is still no clear empirical evidence for a mass-loss bi-stability jump, while a constant Teff of the TAMS is hard to explain with current observational and theoretical constraints on convective overshoot. Note that [20] showed the feature to be present in both the LMC and Milky Way, suggesting the feature is metallicity independent.

While binary mass transfer and merging may well be able to explain (part of) the BSG problem it is by no means certain that this will be the case. It is perhaps somewhat unfortunate that binary merging has been invoked to explain both the *rapid* rotation of B[e] supergiants, as well as the *slow* rotation of the general population of canonical B supergiants.

Large homogeneous samples such as XShooting ULLYSES (XShootU), BLOeM, and WEAVE-SCIP are needed to help resolve the gaps in our understanding of massive star evolution. Moreover, asteroseismology of supergiants will provide promising insights into their interior structures [54–56].

**Author Contributions:** J.S.V. wrote the paper, R.D.O. took care of the Gaia distances. Both authors discussed the scientific contents. All authors have read and agreed to the published version of the manuscript.

**Funding:** JSV acknowledges funding from STFC grant ST/V000233/1 (PI Vink).

**Data Availability Statement:** Data are available upon request.

**Acknowledgments:** We would like to thank Gautham Sabhahit and Erin Higgins for discussion and help with the figures, and Melina Fernandez en Silvina Cárdenas for their help with Gaia data.

**Conflicts of Interest:** The authors declare no conflicts of interest.

## Abbreviations

The following abbreviations are used in this manuscript:

| | |
|---|---|
| B[e] | The B[e] phenomenon refers to forbidden emission lines. |
| BSG | Blue supergiant (not to be confused with B supergiant that refers to spectral type) |
| RSG | Red supergiant |
| LBV | Luminous blue variable |
| CBM | Core-boundary mixing |
| ML-plane | Mass-luminosity plane |
| MS | Main sequence |

## Notes

[1] There is a technical difference in terminology between B supergiants, referring to supergiants of a specific spectral type (i.e., B), and blue supergiants (BSG) which is the more generic evolutionary term that distinguishes the hotter and bluer supergiants (hotter than ∼8 kK) from the cooler (3–5 kK) red supergiants (RSGs).

[2] Note that the reality of the bi-stability jump is under debate [30,31]

[3] Note that this could potentially be improved [37].